\def\bar{\overline}
\def\hat{\widehat}
\def\*{\star}
\def\[{\left[}
\def\]{\right]}
\def\({\left(}
\def\){\right)}

\def\zb{{\bar{z} }}
\def\frac#1#2{{#1 \over #2}}
\def\inv#1{{1 \over #1}}
\def\half{{1 \over 2}}
\def\d{\partial}

\def\vev#1{\langle #1 \rangle}

\def\2pi{\hbox{$2\pi i$}}

\def\dsl{\raise.15ex\hbox{/}\kern-.57em\partial}
\def\Dsl{\,\raise.15ex\hbox{/}\mkern-.13.5mu D}

\def\ga{\gamma}		

\def\al{\alpha}
\def\ep{\epsilon}
\def\la{\lambda}	\def\La{\Lambda}
\def\de{\delta}		
		
\def\sig{\sigma}	

		\def\CO{{\cal O}}

\font\numbers=cmss12
\font\upright=cmu10 scaled\magstep1
\def\stroke{\vrule height8pt width0.4pt depth-0.1pt}
\def\topfleck{\vrule height8pt width0.5pt depth-5.9pt}
\def\botfleck{\vrule height2pt width0.5pt depth0.1pt}
\def\Zmath{\vcenter{\hbox{\numbers\rlap{\rlap{Z}\kern
		0.8pt\topfleck}\kern
		2.2pt \rlap Z\kern 6pt\botfleck\kern 1pt}}}
\def\Qmath{\vcenter{\hbox{\upright\rlap{\rlap{Q}\kern
                   3.8pt\stroke}\phantom{Q}}}}
\def\Nmath{\vcenter{\hbox{\upright\rlap{I}\kern 1.7pt N}}}
\def\Cmath{\vcenter{\hbox{\upright\rlap{\rlap{C}\kern
                   3.8pt\stroke}\phantom{C}}}}
\def\Rmath{\vcenter{\hbox{\upright\rlap{I}\kern 1.7pt R}}}
\def\Z{\ifmmode\Zmath\else$\Zmath$\fi}
\def\Q{\ifmmode\Qmath\else$\Qmath$\fi}
\def\N{\ifmmode\Nmath\else$\Nmath$\fi}
\def\C{\ifmmode\Cmath\else$\Cmath$\fi}
\def\R{\ifmmode\Rmath\else$\Rmath$\fi}
\def\cadremath#1{\vbox{\hrule\hbox{\vrule\kern8pt\vbox{\kern8pt
			\hbox{$\displaystyle #1$}\kern8pt}
			\kern8pt\vrule}\hrule}}

\def\presentation{
\voffset -.50in   
\hoffset -.19in
\oddsidemargin 0in \evensidemargin 0in
\marginparwidth .75in \marginparsep 7pt \topmargin 0in
\headheight 12pt \headsep .25in
\footheight 18pt \footskip .35in
\textheight 9.5in \textwidth 6.5in
\columnsep 10pt \columnseprule 0pt }
\def\debut{ \begin{eqnarray} }
\def\fin{ \end{eqnarray} }
\def\non{ \nonumber }
\documentstyle[12pt]{article}
\presentation
\begin{document}
\centerline{\LARGE The Large N Random Phase sine-Gordon Model.}
\bigskip
\vskip1cm
\centerline{\large  Michel Bauer and Denis Bernard
\footnote[1]{Member of the CNRS} }
\centerline{Service de Physique Th\'eorique de Saclay
\footnote[2]{\it Laboratoire de la Direction des Sciences de la
Mati\`ere du Commisariat \`a l'Energie Atomique.}}
\centerline{F-91191, Gif-sur-Yvette, France.}
\vskip2cm
Abstract.

At large distances and in the low temperature phase, the quenched
correlation functions in the 2d random phase sine-Gordon model
have been argued to be of the form~:
$  \bar {\vev{~[\varphi(x)-\varphi(0)]^2~}}_*
  = A (\log|x|) + B \ep^2 (\log|x|)^2     $,  with $\ep=(T-T_c)$.
However, renormalization group computations predict $B\not=0$
while variational approaches (which are supposed to be exact
for models with a large number of components) give $B=0$.

We introduce a large $N$ version of the random phase sine-Gordon
model. Using non-Abelian bosonization and
renormalization group techniques, we show that the correlation
functions of our models have the above form but with a coefficient
$B$ suppressed by a factor $1/N^3$ compared to $A$.
\vskip 8.0 truecm P.A.C.S.~: 05.70.J; 64.60.F; 64.70.P
\vfill
\newpage
%
\def\bbbox{\begin{picture}(3,3)(-3,-3)
\put(-3,-3){\framebox(3,3)} \end{picture}}

The 2d random phase sine-Gordon model has been introduced
to describe many disordered systems including
the 2d XY model in a random field \cite{Ca}, interfacial
roughening transition \cite{CuSh}, randomly pinned
flux lines in supraconductors \cite{Fis}, etc....
Its action is~:
\debut
S(\varphi|A_\mu;\xi)
= \int \frac{d^2x}{4\pi}\({ \frac{K}{2}
(\d_\nu\varphi)^2 - A_\nu(x) \d_\nu\varphi - \xi(x) e^{i\varphi}
-\xi^*(x) e^{-i\varphi} }\) \label{actA}
\fin
The coupling $K$ is proportional to the inverse temperature,
$K\propto 1/T$. In addition to the random phase, we also
 introduced a random potential
since this is required by one-loop renormalization \cite{Ca}.
The quenched random variables $A_\nu(x)$ and $\xi(x)$ are gaussian
with law
\debut
P[A] = \exp\[{ -\inv{2g} \int \frac{d^2x}{4\pi} A_\mu A_\mu}\]
\quad and\quad
P[\xi] = \exp\[{ -\inv{2\sig} \int \frac{d^2x}{4\pi}~\xi \xi^* }\]
\label{gauss1}
\fin

There are two different phases~:
a low temperature phase $K>K_c=1$ in which the disorder is relevant, and
a high temperature phase $K<K_c=1$ in which the disorder is irrelevant.
At $K=K_c$, the disorder is marginally irrelevant and only induces
logarithmic corrections to the pure system.
Of particular interest is the large distance behavior
of the quenched average of the correlation
functions of $\varphi(x)$. The first proposal for this
correlation in the low temperature phase
was found using renormalization group  (RG)
techniques to be of the following form \cite{Ca,Vil,To}:
\debut
\bar {\vev{~[\varphi(x)-\varphi(0)]^2~}}_* =
A(\log|x|) + B \ep^2 (\log|x|)^2
\label{jungle}
\fin
with $B \neq 0$ and $\ep=\({\frac{K-K_c}{K_c} }\)\ll 1$.
There is a crossover from a $(\log|x|)$ to a $(\log|x|)^2$ behavior.
The crossover length $R_{cross}$ is such that
$\({\log R_{cross} }\)\sim
\frac{A}{B\ep^2}$, for $\ep=\({\frac{K-K_c}{K_c}}\)\ll 1$.
It is exponentially large close to the phase transition.

However the formula (\ref{jungle}) is still controversial,
theoretically as well as numerically. In fact variational
approaches predict $B=0$ and a $(\log |x|)$ behavior
\cite{Kor,Gia,Orl}. The RG flows was also found to be unstable again
assymmetric replica perturbations \cite{Dou}.
The numerical verifications of (\ref{jungle}) are also not settled~:
a $(\log|x|)$ behavior was found in ref.\cite{Bat,Rie}, while more
recent simulations \cite{Mar} seem to indicate a $(\log|x|)^2$ behavior.
In view of this conflict, and since the variational approaches
are argued to be exact for systems with a large number of components
\cite{Mez},
we studied a large $N$ version of the model (\ref{actA}).
As explained below, using  RG
computations based on the  (a priori symmetric) replica trick,
we find that our large $N$ model possesses a non-trivial
infrared fixed point in which the correlation functions have
the form (\ref{jungle}) but with a coefficient $B$ suppressed by a factor
$(1/N^3)$ compared to $A$. The occurence of this factor could explain
why the $(\log|x|)^2$ term does not manifest itself
in the variational approaches.

\bigskip
$\bullet)$ {\it The large N bosonic and fermionic actions.}

To introduce the large $N$ version of (\ref{actA}), it is convenient
to fermionize it. The fermionic form of
the random phase sine-Gordon model is a massless Thirring model
coupled to a quenched potential $A_\mu$ and a random phase $\xi$.
To define its large $N$ version, we need to introduce $N$ Dirac
fermions with components $\psi_\pm^k$ and $\bar \psi_\pm^k$ with
$k=1,\cdots,N$. Let $z=x+iy$ and $\zb= x-iy$ be the complex coordinates
on the plane. The action is~:
\debut
S^{(N)}&=&  \int \frac{d^2x}{\pi}\({ \sum_{k=1}^N
(\psi_{-;k}\d_\zb \psi_+^k + \bar \psi_{-;k}\d_z \bar \psi_+^k)
- \frac{a}{N} (\sum_{k=1}^N \psi_{-;k}\psi_+^k)(\sum_{k=1}^N\bar \psi_{-;k}
\bar \psi_+^k) }\) \label{actferm}\\
&~&- \int \frac{d^2x}{\pi}\({ A_\zb (\sum_{k=1}^N \psi_{-;k}\psi_+^k)
+ A_z (\sum_{k=1}^N\bar \psi_{-;k} \bar \psi_+^k)
+ \xi (\sum_{k=1}^N\bar \psi_{-;k} \psi_+^k)
+ \xi^* (\sum_{k=1}^N\psi_{-;k} \bar\psi_+^k) }\) \non
\fin
In absence of disorder, it is conformally invariant.
The random potential $A_\mu=(A_\zb,A_z)$ is coupled to the
$U(1)$ currents $J_z=\sum_{k=1}^N \psi_{-;k}\psi_+^k$
and $\bar J_\zb=\sum_{k=1}^N\bar \psi_{-;k} \bar \psi_+^k$
of the unperturbed theory. At $\xi=0$, the random potential
does not break conformal invariance.

There are a priori many ways to generalize the action (\ref{actA})
to a large $N$ version. The action (\ref{actferm}) has been
designed in such way as (i) to keep the number of disordered variables
fixed, (ii) to preserve the exact conformal invariance in
absence of disorder, and (iii) to parallel as much as
possible standard properties
of large $N$ models. In particular, we can implement a
Hubbard-Stratonovitch transformation in order to disentangle
the quartic interaction in (\ref{actferm}). The action is then
quadratic in the $N$ fermions and we can integrate over them.
This leads to the following representation of
the partition function at fixed disorder~:
\debut
\int D\psi~ \exp\({-S^{(N)}}\) = \int DQ
\exp\[{-N\({\inv{a}\int \frac{d^2x}{\pi} Q_zQ_\zb
- Tr\({\log H }\) }\) }\] \label{hub}
\fin
with $H=\pmatrix{ \xi & \d_z - A_z - Q_z \cr
		\d_\zb - A_\zb -Q_\zb & \xi^* \cr}$.
Therefore, at large $N$ the path integral (\ref{hub})
is dominated by the saddle point as usual in large $N$ technique.
However, this is not the way we will pursue.

The fermionic action (\ref{actferm}) can be bosonized back using
non-Abelian bosonization \cite{Wit}. As usual, since the pure system describes
$N$ Dirac fermions, the pure bosonized theory will be described by
a $su(N)$ Wess-Zumino-Witten (WZW) model at level one plus a massless
free field. The $su(N)$ WZW model at level one possesses primary
fields taking values in the $(N-1)$ fundamental representations of $su(N)$.
Let $\phi_{\bbbox}^k$ and $\phi_{\bar {\bbbox} ;k}$ be
the chiral WZW primary fields which take values in the
defining representation of $su(N)$ and in its complex conjugate.
Their conformal weights are both equal to $\({\frac{N-1}{2N}}\)$.
Let us denote by $\varphi$ the gaussian free field.
The original fermions $\psi_\pm^k$  can be written as the product
of these WZW primary fields
by a vertex operator of the gaussian model. Namely,
$\psi_+^k= \phi_{\bbbox}^k~e^{\frac{i}{\sqrt{N}}\varphi}$ and
$\psi_{-;k}= \phi_{\bar {\bbbox};k}~ e^{-\frac{i}{\sqrt{N}}\varphi}$,
and similarly for the other chiral components $\bar \psi_+^k$
and $\bar \psi_{-;k}$.
The bosonic form of the action (\ref{actferm}) is~:
\debut
S^{(N)} &=& S_{wzw} + \frac{K}{2}
 \int \frac{d^2x}{4\pi} (\d_\nu\varphi)^2 \non\\
 &~&  - \int \frac{d^2x}{4\pi} \({ A_\nu(x) \d_\nu\varphi +
\xi(x) \Phi ~e^{\frac{i}{\sqrt{N}}\varphi}
+\xi^*(x) \Phi^* ~e^{-\frac{i}{\sqrt{N}}\varphi} }\) \label{actbos}
\fin
where $S_{wzw}$ refers to the $su(N)_1$ WZW action.
We have introduced the composite fields
 $\Phi(z,\zb)=\sum_{k=1}^N \phi_{\bbbox}^k(z)
 \bar {\phi_{\bar {\bbbox};k}}(\zb)$
and $\Phi^*=\sum_{k=1}^N \phi_{\bar {\bbbox};k}(z)
 \bar {\phi_{\bbbox}^k}(\zb)$.
Equivalently, $\Phi=tr_{\bbbox}(G)$ with $G$
the group valued field of the WZW model.
For $N=1$ the WZW terms are absent and we recover the  action
(\ref{actA}) of the random phase sine-Gordon model.
We  used the action (\ref{actbos}) as  definition of the model.

The effect of the Thirring interaction, specified
by the coupling constant $a$, is summarized
in the normalization of the gaussian action for $\varphi$.
That is, $K$ is a known function of $a$ whose explicit form is
not needed. The coefficient $K$ fixes the scale of
the dimension of the vertex operators~:
$\dim \({ e^{\frac{i}{\sqrt{N}}\varphi} }\)= \inv{KN}$.
Since the dimension of $\Phi$ is $\({\frac{N-1}{N}}\)$,
the dimension of the field coupled to the random variable $\xi$
is~:
\debut
h \equiv
\dim\({ \Phi_{\bbbox}~e^{\frac{i}{\sqrt{N}}\varphi} }\)
= \frac{N-1}{N}+ \inv{KN} = 1+ \({\frac{1-K}{KN}}\). \label{dimpert}
\fin
{}From the Harris criterion we know that the disordered
 variable $\xi$
will be relevant if this dimension is less than one.
Thus, the critical temperature is $K_c=1$~:
 with our convention the critical temperature is
independent of $N$. For $K>K_c$ the disorder is relevant while
for $K<K_c$ it is irrelevant.

The WZW model possesses chiral conserved currents, which
we denote by $J_z^a(z)$ and $\bar {J_\zb}^a(\zb)$ with
$a=1,\cdots,\dim su(N)$. In order to fixe
the normalization in the WZW sector
for later convenience, we  give the
operator product expansions of the primary fields $\phi_{\bbbox}$
and $\phi_{\bar {\bbbox}}$ and of the currents~:
\debut
J^a_z(z) J^a_z(0) &=& \frac{I_{\bbbox}\delta^{ab}}{z^2}
+\frac{if^{abc}}{z}~J^c_z(0) +\cdots \label{ope1}\\
J^a_z(z) \phi_{\bbbox}^k(0) &=&\frac{ t^{a~k}_{~j}}{z}~
\phi_{\bbbox}^j(0) +\cdots\label{ope2}\\
\phi_{\bbbox}^k(z) \phi_{\bar {\bbbox};j}(0)
&=& \inv{z^{\frac{N-1}{N}}} \({ \delta^k_j
+ z~ I_{\bbbox}^{-1}~t^{a~k}_{~j}~J^a_z(0) }\)
+\cdots \label{ope3}
\fin
where $[t^a,t^b]=-if^{abc}t^c$ with $f^{abc}$ the structure
constants of $su(N)$ and $tr(t^at^b)=I_{\bbbox} \delta^{ab}$.
Eq.(\ref{ope1}) encodes the fact that the level of the
representation of the $su(N)$ affine algebra  is one.
We will also need the Casimir operator whose value in the adjoint
representation is denoted by $C_G$ and in the defining
representation by $C_{\bbbox}$. We have $NC_{\bbbox}=I_{\bbbox} d_G$
with $d_G=\dim (su(N))= N^2-1$. A convenient normalization
is $C_G=2N$, $C_{\bbbox}=(N^2-1)/N$ and thus $I_{\bbbox}=1$.

Similarly as the random phase sine-Gordon model \cite{Bre},
the large $N$ model  possesses a $U(1)$ symmetry which
amounts to absorbe any translation of
$\varphi$ into the random variables $A_\mu$ and $\xi$.
Its Noether current is represented by insertions of
$(\d_\mu\varphi)$ in the quenched connected correlation functions.
This symmetry  implies that all the quenched
connected correlation functions of $(\d_\mu\varphi)$
are unaffected by the disorder. But
this symmetry has another consequence~: the $g$-dependence
of the quenched correlation functions can be factorized.
The simplest way to visualize it consists in
noticing that the $A_\mu$-dependence of the action
can be absorbed into a shift of $\varphi$.
Indeed, let us decompose $A_\mu$ as
$A_\mu=\d_\mu \La + \ep_{\mu\nu}\d_\nu\zeta$, (this is
always possible on the plane). The  field $\zeta$
decouples from the action and from the measure. It is
therefore irrelevant and we can set it to zero.
Denoting by $S^{(N)}(\varphi|\La;\xi)$ the action
(\ref{actbos}) with $A_\mu=\d_\mu \La$, we have~:
\debut
S^{(N)}(\varphi|\La,\xi) = S^{(N)}(\varphi-\La/K|\La=0,\hat \xi)
-\inv{2K} \int \frac{d^2x}{4\pi} (\d_\mu(\La))^2, \non
\fin
with $\hat \xi= \xi e^{i\La/K}$.
For the correlation functions involving the vertex operators
$e^{i\al_p\varphi(x_p)}$,  this implies~:
\debut
\vev{ e^{i\al_1\varphi(x_1)}\cdots}_{\La,\xi}
= \({\prod_p e^{i\frac{\al_p}{K}\La(x_p)}}\)~
\vev{ e^{i\al_1\varphi(x_1)}\cdots}_{\La=0,\hat \xi} \label{shiftbis}
\fin
Let $G_{\al_1,\cdots}(x_1,\cdots|g,\sig)$ be their quenched averages.
Using the fact that $\hat \xi$ and $\xi$ have the same measure, and
integrating (\ref{shiftbis}) over $\La$ using the free field gaussian
measure (\ref{gauss1}), $P[A_\mu]=
\exp\[{ -\inv{2g} \int \frac{d^2x}{4\pi} (\d_\mu\La)^2 }\]$ with
$A_\mu=\d_\mu \La$, we deduce that~:
\debut
G_{\al_1,\cdots}(x_1,\cdots|g,\sig)
=\prod_{p<q}|x_p-x_q|^{2g\al_p\al_q/K^2}~~
G_{\al_1,\cdots}(x_1,\cdots|g=0,\sig) \label{Gg}
\fin
Equivalently,
\debut
\d_g G_{\al_1,\cdots}(x_1,\cdots|g,\sig)
=\({\sum_{p<q} \frac{\al_p\al_q}{K^2} \log(|x_p-x_q|^2)}\)~
G_{\al_1,\cdots}(x_1,\cdots|g,\sig) \label{gdepend}
\fin
This Ward identity will be useful for analyzing the
renormalization group equations.

\bigskip
$\bullet$) {\it The effective action and the beta functions.}

To compute the renormalization group equations we used the
(a priori symmetric) replica trick.
Introducing $n$ copies
of the system with the same disorder and then averaging over the
disorder gives the following effective action~:
\debut
S_{eff}^{(N)}&=&  \sum_r S_{wzw}^{(r)} +\half
\int \frac{d^2x}{4\pi} (\d_\mu\varphi^r) G_{rs}(\d_\mu\varphi^s) \non\\
 &~&- 2\sig \int \frac{d^2x}{4\pi}
\sum_{r\not= s} \({ \Phi^r~\Phi^{*~s}~
e^{\frac{i}{\sqrt{N}}(\varphi^r-\varphi^s)} }\)
 - \la \int \frac{d^2x}{4\pi}  \sum_r
\({\sum_a J^{a;r}_z \bar {J_\zb}^{a;r} }\)
\label{actrsg}
\fin
with $G_{rs}= K\de_{rs}-g$.
In eq. (\ref{actrsg}), the indices $r,s,\cdots$ refer to the
replica index and therefore run from $1$ to $n$.
The fields with a replica index refer to the copies
of the original fields in the $r^{th}$ replicated system.
The extra $\la$-term arises from the regularization of
the $\sig$-term for $r=s$. Even if we would not have introduced
it at this point, it would have been generated at one-loop.
Since it does not couple the replica, we could
have introduced it  in eq.(\ref{actbos}). There it
represents a current-current interaction in the WZW sector.

Since the interaction only involves the difference of
the fields $\varphi^r$, it is convenient to
decompose their kinetic term as follows~:
\debut
\half  (\d_\mu\varphi^r) G_{rs}(\d_\mu\varphi^s)
= \frac{(K-ng)}{2n}\({\d_\mu (\sum_r\varphi^r)}\)^2
+ \frac{K}{4n} \sum_{r\not= s} \({\d_\mu(\varphi^r-\varphi^s)}\)^2
\label{kine}
\fin
The field $\({\sum_r\d_\mu\varphi^r}\)$ decouples.
Since its correlation functions
represent the averages of the connected correlation
functions of the $U(1)$ current, we recover  that
these correlations are unaffected by the disorder.
The Ward identity (\ref{gdepend}) can also be recovered
using this decomposition.

The renormalization group allows us to perturbatively
analyse the behavior of the system in the low temperature
phase. We present a one-loop computation, which is
valid close to the critical temperature,
ie. $\({\frac{K-K_c}{K_c}}\)\ll 1$.
We recall that if the partition function and correlation functions
are computed with the measure $\int D\phi ~ \exp(-S)$
with $S = S_* + \sum_i g^i \int d^2x \CO_i(x)$,
where $S_*$ is the unperturbed fixed point action and
$\CO_i(x)$ are a set of relevant primary operators
of dimension $h_i$, the beta functions at one-loop are~:
\debut
\dot g^i = \beta^i(g) =
(2-h_i) g^i - \pi \sum_{jk}C^i_{jk} g^j g^k + \cdots
\label{A3}
\fin
The summation in the second term
is over all the relevant fields generated by the
 operator product expansions, the coefficients $C^i_{jk}$ are
determined by the relation~:
$\CO_i(x)\CO_j(0) = \sum_k |x|^{h_k-h_i-h_j} C_{ij}^k \CO_k(0) + \cdots$

Let us introduce the following notation for the perturbing
fields in the action (\ref{actrsg})~:
\debut
\CO_1 &=& \sum_{r\not= s} \({\Phi^r~\Phi^{*~s}~
e^{\frac{i}{\sqrt{N}} (\varphi^r-\varphi^s)} }\) \label{O1}\\
\CO_2 &=& \inv{2n} \sum_{r,s} i\d_z(\varphi^r-\varphi^s)
i\d_\zb(\varphi^r-\varphi^s)
\label{O2}\\
\CO_3 &=& \sum_r \({\sum_a J^{a;r}_z \bar {J_\zb}^{a;r} }\) \label{O3}
\fin
The fields $\CO_1$ and $\CO_3$ are the perturbing fields associated to the
coupling constants $\sig$ and $\la$, respectively.
The field $\CO_2$ is one of the kinetic field for the $\varphi^r$'s.
We have the following operator product expansions~:
\debut
\CO_1(z)\CO_1(0) &=& \frac{2N(n-2)}{|z|^{2h}} \CO_1(0)
+ \frac{ 2Nn}{|z|^{2h-2}} \CO_2(0)
- \frac{2N(n-1)/I_{\bbbox}}{|z|^{2h-2}} \CO_3(0) + \cdots \non\\
\CO_3(z)\CO_3(0) &=& -\frac{C_G}{|z|^2} \CO_3(0) + \cdots\label{opesg}\\
\CO_3(z)\CO_1(0) &=& -\frac{2C_{\bbbox}}{|z|^2} \CO_1(0) + \cdots\non\\
\CO_2(z)\CO_1(0) &=& \frac{2/NK^2}{|z|^2} \CO_1(0)
+ \cdots \non
\fin
where the dots refer to irrelevant terms.
The operator product expansions $\CO_2(z)\CO_2(0)$ and
$\CO_3(z)\CO_2(0)$ only contain  irrelevant terms.
Using the formula (\ref{A3}) we get the beta functions at
finite $n$.
Since the field $(\sum_r\varphi^r)$ decouples, the coupling $(K-ng)$ is
unrenormalized at any order in perturbation theory, and
$\beta_K = n \beta_g$.
Setting $n=0$ as required by the replica trick, we get
for $\({\frac{K-K_c}{K_c}}\)\ll 1$~:
\debut
\beta_\sig &=& \frac{2}{N}\({\frac{K-K_c}{K_c
}~}\) \sig - 2N\sig^2
- C_{\bbbox}\la\sig+\cdots \non\\
\beta_\la &=& -\({ \frac{C_G}{4} \la^2
- \frac{2N}{I_{\bbbox}}\sig^2}\) +\cdots \label{betasg}\\
\beta_g &=& \frac{N}{2}\sig^2 +\cdots\non\
\fin
and $\beta_K =0$. Here the dots refer to higher loop contributions.
So, $K$ is unrenormalized at $n=0$.
These equations show that the couplings to $A_\mu$ and to $\CO_3$ are
generated at one-loop even if one does not include them at tree level.
Notice the opposite signs in the $\la^2$ and $\sig^2$ contribution
to $\beta_\la$.

  From equation (\ref{betasg}), we immediatly see
that in the low temperature phase $(K>K_c)$, the beta functions
$\beta_\sig$ and $\beta_\la$ possess a non-trivial infrared zero at
$\sig_*$ and $\la_*$ given by~:
\debut
\la_*^2 &=& 8\({ \frac{d_G}{C_{\bbbox} C_G} }\)~\sig^2_* \non\\
 \frac{2}{N} \({\frac{K-K_c}{K_c}}\) &=&
2N \sig_*+ C_{\bbbox}~\la_*
\sim_{N\to\infty} 4N ~\sig_*
\label{IRpoint}
\fin
The last beta function $\beta_g$ does not
vanish at $\sig_*,~\la_*$, but it behaves like~:
\debut
\beta_g^* = \inv{8N^3} \({\frac{K-K_c}{K_c}}\)^2 +\cdots,
\quad for\quad N\gg 1,\quad \frac{K-K_c}{K_c}\ll 1
\label{betastar}
\fin
Hence, even at $\sig=\sig_*$ and $\la=\la_*$, the coupling $g$
will continue to flow. We may characterize
such pseudo-fixed point as a ``run away fixed point".

\bigskip
$\bullet$) {\it The RG equations and the
quenched correlation functions.}

The non-vanishing of the beta function $\beta_g$ at the
infrared pseudo-fixed point has direct consequences.
Consider the quenched correlation  functions
$G_{\al_1,\cdots}(x_1,\cdots)$ involving the vertex operators
$e^{i\al_p\varphi(x_p)}$.
They satisfy the renormalization group equations~:
\debut
\[{ \sum_p x_p^\nu\frac{\d}{\d x_p^\nu} + \sum_p \ga_p -
\sum_{c=g,\sig,\la}\beta_{c}\frac{\d}{\d c} }\]
G_{\al_1,\cdots}(x_1,\cdots)=0 \non
\fin
Here $\ga_p$ are the anomalous dimensions. Since the $g$-dependence is
explicitly known from eq.(\ref{gdepend}), we can separate the function
$\beta_g$ from $\beta_\sig$ and $\beta_\la$ to get~:
\debut
\[{ \sum_p x_p^\nu\frac{\d}{\d x_p^\nu} + \sum_p \ga_p -
\sum_{c=\sig,\la}\beta_{c}\frac{\d}{\d c} -
\beta_g \sum_{p<q} \frac{\al_p\al_q}{K^2}\log(|x_p-x_q|^2)}\]
G_{\al_1,\cdots}(x_1,\cdots)=0 \non
\fin
In particular at the infrared fixed
point $\sig_*$, $\la_*$, in which $\beta_\sig^* =\beta_\la^*=0$, we get~:
\debut
\[{ \sum_p x_p^\nu\frac{\d}{\d x_p^\nu} + \sum_p \ga_p^* -
\beta_g^* \sum_{p<q} \frac{\al_p\al_q}{K^2}\log(|x_p-x_q|^2)}\]
G_{\al_1,\cdots}^*(x_1,\cdots)=0 \label{rgeq2}
\fin
where $\ga_p^*$ are the values of the anomalous dimensions
at the infrared fixed point and
$\beta_g^*$ is given in eq.(\ref{betastar}) for large $N$ and
for $\({\frac{K-K_c}{K_c}}\)\ll 1$.
The net effect of the $g$-flow is to add the extra logarithmic term in
the renormalization group equations (\ref{rgeq2}).

Eq.(\ref{rgeq2}) can be used to compute two-point functions at the infrared
fixed point. Consider~:
\debut
G_1(x)&=& \bar { \vev{\exp\({i\al(\varphi(x)-\varphi(0)}\)}  }_* \non\\
G_2(x)&=& \bar {\vev{\exp\({i\al\varphi(x)}\)}~
\vev{\exp\({-i\al\varphi(0)}\)}}_*\non
\fin
Let $\ga_{1,2}^*$ be their anomalous dimensions at the infrared fixed point.
The renormalization group equation (\ref{rgeq2}) gives~:
\debut
G_{1,2}(x)= |x|^{-2\ga_{1,2}^*}~
\exp\({-\frac{ \al^2\beta_g^* }{2K^2} (\log|x|)^2 }\)
\label{corrlog}
\fin
Notice the $(\log|x|)^2$ correction which arises from the renormalization
of $g$. The anomalous dimensions are even in the charge $\al$ and vanishes
at $\al=0$, therefore~:
\debut
 \ga_{1,2}^* = \frac{\al^2}{K} \rho_{1,2}^* + \CO(\al^4),
\quad with\quad \rho_{1}^*=1 +\CO\({\frac{K-K_c}{K_c}}\)
\label{dimir}
\fin
Expanding eq.(\ref{corrlog}) in power of $\al^2$
gives the two-point functions of $\varphi$~:
\debut
 \bar {\vev{~[\varphi(x)-\varphi(0)]^2~}}_* =
\frac{2\rho_{1}^*}{K}\log|x| + \frac{\beta_g^*}{2K^2} (\log|x|)^2
\label{ver1}\\
 \bar {\[{~ \vev{\varphi(x)-\varphi(0)}~ }\]^2 }_*=
\frac{2\rho_{2}^*}{K}\log|x| + \frac{\beta_g^*}{2K^2} (\log|x|)^2
\label{ver2}
\fin

Comparing the formula for $\rho_{1}^*$ and $\beta_g^*$ close
to the phase transition, eq. (\ref{dimir},\ref{betastar}), we see that
at large $N$ the $(\log |x|)^2$ term is suppressed by a factor of
$1/N^3$ compared to the $(\log |x|)$ term.
Hence there is no a priori contradiction between the replica
symmetric RG computations and the variational approaches~:
they coincide in the infinite $N$ limit, i.e. in the regime where the
latter are expected to be exact. It would be interesting to compute
the $\inv{N}$ corrections in the variational approaches for comparison.

Note that the $(\log|x|)^2$ cancel in the connected correlation
function $\bar {\vev{[\varphi(x)-\varphi(0)]^2}_{conn}}$ as it should be,
since this connected correlation function is unaffected by the disorder.
Eq. (\ref{ver1}, \ref{ver2}) are exact to all orders in perturbation theory
provided we may trust the renormalization group using symmetric
replica. Only the exact value of the beta function $\beta^*_g$
depends on the order of the perturbation theory.

{\it Acknowledgement~:} It is a pleasure to thank Henri Orland
for discussions.

\end{document}